\documentclass[sigconf]{acmart}

\AtBeginDocument{%
	\providecommand\BibTeX{{%
			\normalfont B\kern-0.5em{\scshape i\kern-0.25em b}\kern-0.8em\TeX}}}

\copyrightyear{2021}
\acmYear{2021}
\setcopyright{acmlicensed}
\acmConference[CIKM '21]{Proceedings of the 30th ACM International Conference on Information and Knowledge Management}{November 1--5, 2021}{Virtual Event, QLD, Australia}
\acmBooktitle{Proceedings of the 30th ACM International Conference on Information and Knowledge Management (CIKM'21), November 1--5, 2021, Virtual Event, QLD, Australia}
\acmPrice{15.00}
\acmDOI{10.1145/3459637.3482257}
\acmISBN{978-1-4503-8446-9/21/11}
\usepackage{multirow}
\usepackage{graphicx}
\usepackage{float} 
\usepackage{subfigure}
\usepackage{threeparttable}
\usepackage{enumitem}
\usepackage{dutchcal}
\usepackage{adjustbox}
\newcommand{\tabincell}[2]{\begin{tabular}{@{}#1@{}}#2\end{tabular}}  
\usepackage[T1]{fontenc}
\usepackage[skip=1ex]{caption}
\usepackage{booktabs,tabularx}
\usepackage{siunitx}
\newcolumntype{C}{>{\centering\arraybackslash}X}

\begin{document}
	\title{Extracting Attentive Social Temporal Excitation for Sequential Recommendation}
	\author{Yunzhe Li$^{1}$, Yue Ding$^{1}$, Bo Chen$^{2}$, Xin Xin$^{3}$} \author{Yule Wang$^{1}$, Yuxiang	Shi$^{1}$,Ruiming Tang$^{2}$, Dong Wang$^{1\dagger}$} 
	\affiliation{$^{1}$Shanghai Jiao Tong Unversity, $^{2}$Huawei Noah's Ark Lab, $^{3}$University of Glasgow, $\dagger$~Corresponding authors}
	\email{{liyzh28, dingyue,wyl666, shiyuxiang, wangdong}@sjtu.edu.cn} \email{{chenbo116, tangruiming}@huawei.com, x.xin.1@research.gla.ac.uk}
\fancyhead{}

	\begin{abstract}
		In collaborative filtering, it is an important way to make full use of social information to improve the recommendation quality, which has been proved to be effective because user behavior will be affected by her friends. However, existing works leverage the social relationship to aggregate user features from friends' historical behavior sequences in a user-level \textit{indirect paradigm}. A significant defect of the indirect paradigm is that it ignores the temporal relationships between behavior events across users. In this paper, we propose a novel time-aware sequential recommendation framework called Social Temporal Excitation Networks (STEN), which introduces temporal point processes to model the fine-grained impact of friends' behaviors on the user's dynamic interests in an event-level \textit{direct paradigm}. Moreover, we propose to decompose the temporal effect in sequential recommendation into social mutual temporal effect and ego temporal effect. Specifically, we employ a social heterogeneous graph embedding layer to refine user representation via structural information. To enhance temporal information propagation, STEN directly extracts the fine-grained temporal mutual influence of friends' behaviors through the \textit{mutually exciting temporal network}. Besides, user's dynamic interests are captured through the \textit{self-exciting temporal network}. Extensive experiments on three real-world datasets show that STEN outperforms state-of-the-art baseline methods. Moreover, STEN provides event-level recommendation explainability, which is also illustrated experimentally.
	\end{abstract}
	
	\maketitle
	
	\section{Introduction}
	In recent years, sequential recommendation, as an emerging topic for capturing user dynamic interests from user behaviors, has received extensive attention from the industry and academia. Various methods emerge to utilize deep sequential models, such as Recurrent Neural Networks (RNNs) \cite{DBLP:journals/corr/HidasiKBT15}, Convolution Neural Networks (CNNs) \cite{2018Personalized}, Attention mechanisms \cite{li2017neural,kang2018self} to extract features from the successive behavior event sequence and enhance recommendation performance. Moreover, some works \cite{bai2019ctrec,li2020time} model the time interval to capture users' demands over time. Nevertheless, for an enormous set of items, only a tiny fraction is involved in a single user behavior sequence. This phenomenon limits the performance of sequential recommendation. 
	
	Furthermore, based on the social influence and social homophily \cite{Resnick1997Recommender, Marsden1994Network, McPherson2001BIRDS}, social relationship plays an important role in recommending suitable items for users. Some works leverage the social relations to aggregate similar users' preferences to enhance collaborative filtering and matrix factorization or integrate deep neural networks into probabilistic matrix factorization for rating prediction \cite{Tang2013Exploiting, zhao2014leveraging,2017Collaborative,2011Recommender,fan2018deep}. Moreover, Graph Neural Networks (GNNs) are also utilized to enhance the performance of recommendation for better capturing the structural information of social relations \cite{2019Graph,mu2019graph,wu2019neural,wu2019dual}. However, these approaches are coarse-grained that only model user-level relationships and ignore the influence of the user behavior sequence. 
	
\begin{figure}
		\setlength{\belowcaptionskip}{-0.5cm} 
		\setlength{\abovecaptionskip}{-0cm} 
		\centering
		\includegraphics[width=0.45\textwidth]{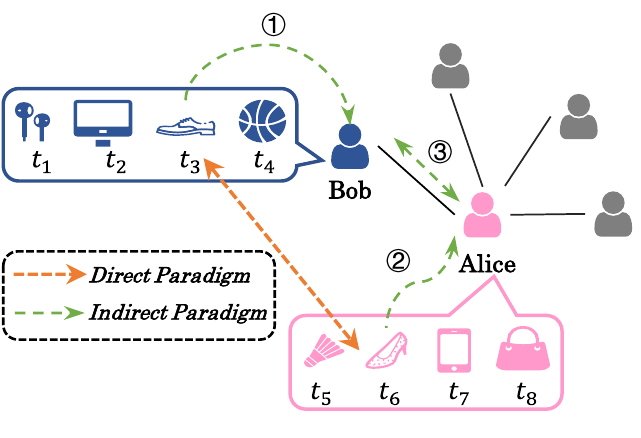}
		\caption{An illustration of two paradigms to deal with asynchronous behavior event sequences and social relations in online communities.}
		\label{fig:socialrec}
	\end{figure}

	To leverage social influence to enhance sequential recommendation performance, existing methods \cite{sun2018attentive,song2019session} usually model both social relations and user's behavior sequences in an indirect manner. As Figure \ref{fig:socialrec} shows, the \textit{indirect paradigm} of the methods (the green path) usually can be summarized in three steps: (1) learning friends' representation from their behavior sequences; (2) learning the target user's representation from her own behavior sequence; (3) aggregating the influence of friends on the target user through their representations. However, the indirect paradigm usually suffers from information loss due to the ignorance of the relationships between behavior events across users. Compared with the indirect paradigm, modeling the influence of different user behavior sequences in the \textit{direct paradigm} (the orange path) is more granular to avoid the problem mentioned above and better model the interaction between cross-user events. 

	Nonetheless, to the best of our knowledge, existing works do not contemplate such a direct modeling paradigm, leaving vast scope for exploration.
	Besides, unlike the synchronous sequence taken at consecutive equal intervals, the users' behavior sequences tend to be asynchronous when considering interactions across user events. Different from only taking the order of events into consideration, how to leverage the temporal information is another challenge for properly modeling the asynchronous sequences.  
	We use a simple example shown in Figure \ref{fig:socialrec} to illustrate this issue. If \textit{Alice} buys heels within a short time after her friend \textit{Bob} buys dress shoes, \textit{i.e.}, $t_3\leq t_6$, then it indicates that \textit{Bob's} behavior is more likely to affect her (\textit{e.g.}, attending certain common activities).
	But if $t_3\ll t_6$ or $t_3>t_6$, \textit{Bob's} behavior will have less or no effect on \textit{Alice}. If the temporal information is lost, the model will not be able to determine the specific relationship between the two events. Hence, the time interval should be treated as important as the order, especially when we note the interaction of cross-user events. 
	
	To address the problems mentioned above, we propose a novel approach called \textbf{S}ocial \textbf{T}emporal \textbf{E}xcitation \textbf{N}etworks (\textbf{STEN}), which models individual user's sequence and the interaction between users together. STEN is designed to learn the influence of temporal information by modeling the temporal effect, which is decomposed into social mutual temporal effect and ego temporal effect. Specifically, the former aggregates the impact of friends' behavior sequences over time, and the latter describes the changes in interests reflected by the user’s own behavior sequence.
	Our model introduces the temporal point process, which is a powerful tool for asynchronous sequence modeling, to model the effect of related cross-user events with temporal information in a fine-grained manner. Specifically, node representation of both users and items are extracted from the social relationship structure and interactions by a \textit{social heterogeneous graph embedding layer}. To explore the impact of cross-user events in a direct paradigm, we model the interaction of friends' behavior event sequences as mutually exciting processes and enhance it by attention mechanism to better describe the social mutual temporal effect in the \textit{mutually exciting temporal network}. Finally, a \textit{self-exciting temporal network} is employed to model the target user's ego temporal effect and generate the user's temporal representation. To summarize, we make the following contributions in this paper: 
	
	\begin{itemize}[leftmargin=*]
		\item  We propose to take temporal information and social relationships into account in the sequential recommendation, \textit{i.e.}, model the impact of friends’ behavior event sequences at the event level. To the best of our knowledge, it is the first attempt to enhance sequential recommendation performance with event-level social information in a direct paradigm.
		\item We propose a novel time-aware sequential recommendation model  STEN. Two temporal excitation networks are designed for modeling both the event-level dynamic influence of social relationship and user's dynamic interests. In particular, STEN can model the fine-grained associations of cross-user events.
		\item We conduct extensive experiments on three real-world datasets and the results demonstrate that STEN outperforms state-of-the-art recommendation models and the effectiveness of the modules in STEN. Besides, we illustrate the advantage of the event-level explainability of our model.  
	\end{itemize}
	

	\begin{figure*}
		\centering
		\includegraphics[width=1\textwidth]{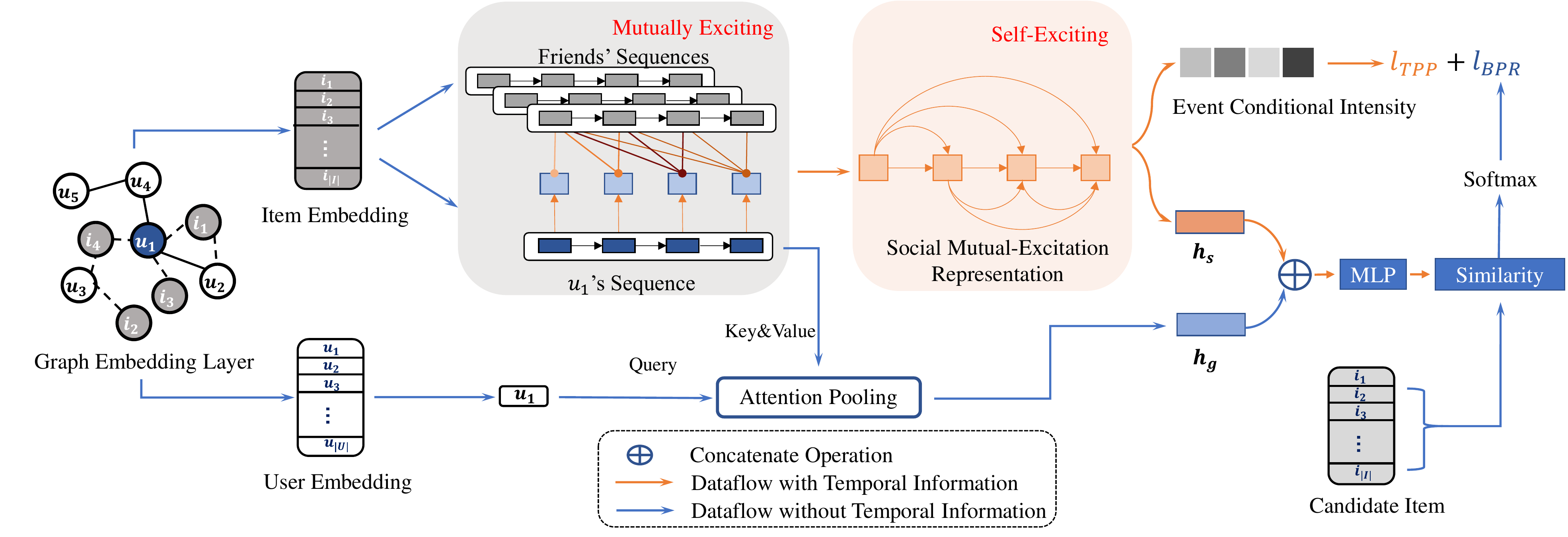}
		\caption{The architecture of STEN. Firstly, through the social graph embedding layer, the node representation of users and items is generated by leveraging the sequence of events and social relations. Then, the embeddings of events are sent to the mutually exciting attention networks to obtain the social mutual temporal representations.  Next, mutual temporal representations are fused in self-exciting temporal networks to capture the temporal dynamic representation $h_s$. Besides, vanilla attention is adopted to get the general interests $h_g$. After that, $h_s,h_g$ are fused to make recommendation. In addition, here adopts joint learning for both fitting the temporal point process and recommendation. }
		\label{fig:gf}
		\vspace{-0.4cm}
	\end{figure*}
	
	\section{Preliminary}
	In this section, we first introduce the temporal point process as background and then give the problem formulation. 
	
	\subsection{Temporal Point Process}
	A temporal point process is a stochastic process consisting of a series of events in the continuous-time domain. Let $ H_t=\{t_i|t_i<t\}$ be the set of historical events that happened before time $t$, where $\{t_1,t_2,...,t_n\}$ is the time sequence of past events. To characterize the Point Process $\mathcal{P}(\lambda(t|H_t))$ for arbitrary time $t$ in a certain time interval $[0,T]$, a conditional intensity function $\lambda(t|H_t)$ is imported to represent the intensity function of the event occurring at time $t$ under the condition of the subset of history events $H_t$ up to time $t$.
	Based on the conditional intensity function and the past events in $\{t_1,t_2,...,t_n\}$, the probability density function of the next event at time $t_{n+1}$ is obtained as follows.
	\begin{equation}
		p(t_{n+1}| t_1,t_2,...,t_n) = \lambda(t_{n+1}|H_{t_{n+1}})exp\Big \{-\int_{t_n}^{t_{n+1}}\lambda(t|H_t)dt\Big \}.
	\end{equation} 
	The probability density function of a whole observed event sequence occurred at $\{t_1,t_2,...,t_n\}$ in interval $[0,T]$ could be represented by:
	\begin{equation}
		\label{eq:rtpp}
		p(\{t_1,t_2,...,t_n\}) = \prod_{i = 1}^{n} \lambda(t_i|H_{t_i}) exp\Big \{-\int_{0}^{T}\lambda(t|H_t)dt\Big \}.
	\end{equation}
	The log-likelihood function $\mathscr{l}$ of the observing event sequence is: 
	\begin{equation}
		\label{eq:lrtpp}
		\mathscr{l}(p(\{t_1,t_2,...,t_n\})) = \sum_{i = 1}^{n} log(\lambda(t_i|H_{t_i})) -\int_{0}^{T}\lambda(t|H_t)dt.
	\end{equation}
	Denote as Eq. (\ref{eq:lrtpp}), the conditional intensity function plays an essential role in mapping the history event sequence to the probability density function. Take the Hawkes Process \cite{10.1093/biomet/58.1.83} as an example, the conditional intensity function of this popular point process is denoted as Eq. (\ref{eq:hpi}):
	\begin{equation}
		\lambda(t|H_t) =\sum_{t>t_i} \varphi(t-t_i),
		\label{eq:hpi}
	\end{equation}
	
	where $\varphi(\cdot)$ is a non-negative function. Eq. (\ref{eq:hpi}) shows that the historical events have an incentive effect on the occurrence of future events and the influence of historical events is superimposed in the form of accumulation. Based on whether there are interactions between multiple processes, the Hawkes Process could be divided into mutually exciting and self-exciting processes.
	
	
	%
	\subsection{Problem Formulation}
	The time-aware sequential recommendation with social information takes both user behavior sequences and their social relations into consideration. Let $U$, $I$ denote the set of users and items respectively and $|U|, |I|$ represent the total number of users and items. For each user $u\in U$, her behavior sequence is described as $\mathbf{s_u}=\{i^u_{t_1},i^u_{t_2},...,i^u_{t_l}\}$, where $l$ is the length of sequence and $i^u_{t_j}\in I$ denotes the item that user $u$ has interacted at time $t_j$. The set of all users' behavior sequences is defined as $S = \{\mathbf{s_{1}},\mathbf{s_{2}},...,\mathbf{s_{|U|}}\}$. For a subset of users $F\subset U$, the set of their behavior sequences is denoted as $\mathbf{S_F}$.
	In addition, user's social relationship could be defined as a social network $G = (V,E)$, where $V$ denotes the node set of users and $E$ is the edge set that represents social links between users. Formally, given users' social relations network $G$ and their behavior sequences $S$, our task is to make a prediction and recommend the next item that the user may be interested in.

	\section{METHODOLOGIES}
	As mentioned above, to take the influence of both behavior sequences and social relations into consideration, we design a novel time-aware sequential recommendation model STEN, which introduces temporal point processes to directly model the fine-grained impact of friends' behaviors at the event level. 
	\vspace{-0.2cm}
	\subsection{Model Architecture}
	The architecture of our model STEN is shown in Figure \ref{fig:gf}. Specifically, we leverage social relations and user behavior sequences to generate the social node embedding for both users and items through the social graph embedding layer. To capture the ordinary social demands, vanilla attention is adopted to get the general interests of the target user. To directly model the impact of friends' behavior sequences, the embeddings of their interacted sequences are sent to the mutually exciting temporal network as mutually exciting processes to generate the social mutual temporal representations. To establish the connection within the sequences, these social mutual temporal representations are fused in the self-exciting temporal network to capture the target user's dynamic interests. After that, we fuse the general interests and the dynamic interests to make recommendations. To guarantee effective learning from the temporal point process, we adopt joint learning in the training stage for both fitting the temporal point process and recommendation.
	\subsubsection{Social Heterogeneous Graph Embedding}
	The social graph embedding aims to extract the general information from the social relations and interactions. In our model, we leverage a heterogeneous graph to represent the relationships for better scalability. For a social heterogeneous network $G_H=(V_H, E_H)$, $\mathit{V_H}$ denotes the nodes of both items and users, $E_H$ denotes the edge set which involves two types of edges: (1)  user-user social links; (2) user-item interaction connection. We first generate the embedding matrices $U\in \mathbb{R}^{|U|\times d}$, $I\in \mathbb{R}^{|I|\times d}$ via embedding lookup, where $d$ denotes the dimensionality of embedding. To tackle the two types of edge, we give two kinds of aggregation to model both users and items. For users, we could directly establish the social relation network through user-user edges and then apply attention aggregation on it, shown as Eq.(\ref{eq:ugat}):
	
	\begin{equation}
		h_u = \sigma(\sum_{j\in N_u}a_{uj}\mathbf{W}_Uh_j+b_U),
		\label{eq:ugat}
	\end{equation}
	where $h_u\in \mathbb{R}^d$ is latent feature of user $u$, $\sigma(\cdot)$ is the non-linear activation function, $N_u$ denotes the neighbor set of user $u$, $\mathbf{W}_U\in \mathbb{R}^{d\times d}$ and $b_U\in \mathbb{R}^d$ are weights and bias of the graph network, $a_{uj}$ is the attention coefficient of user $u$ to user $j$. The attention coefficient is obtained by applying a similarity calculation: 
	\begin{equation}
		a_{uj} = \cfrac{exp(\sigma(v_u[\mathbf{W}_Uh_u||\mathbf{W}_Uh_j]))}{\sum_{k\in N_u} exp(\sigma(v_u[\mathbf{W_U}h_u||\mathbf{W}_Uh_k]))},
		\label{eq:ac}
	\end{equation}
	where $||$ denotes the concatenation operation between two vectors and $v_u\in \mathbb{R}^{2d}$ is the weight vector for users.
	
	%

	Different from user links, there is no direct relationship between items. To establish the relation, we leverage meta-path to find high-order neighbors around the heterogeneous graph\cite{wang2019heterogeneous}. If two items are interacted with the same user, they are more likely to have something in common that satisfies user's interests. Moreover, the items interacted by the same social group are more likely to have relationship. Based on the assumption, we define the $k$-order social meta-path $\phi_k$ as $\{i_m,u_1,u_2,...,u_k,i_n\}$, which denotes a combination relation between item $i_m$ and $i_n$, $k$ is the number of users between two items.
	$N^{\phi_k}_i$ denotes the $k$-order meta-path neighbors of item $i$. In the next, the attention aggregation is applied to the item vertices, shown as follows: 
	
	\begin{equation}
		h^{\phi_k}_i = \sigma(\sum_{j\in N^{\phi_k}_i}a_{ij}\mathbf{W}_Ih_j+b_I),
		\label{eq:igat}
	\end{equation}
	where $h^{\phi_k}_i$ is the $k$-order representation of item $i$, $\mathbf{W}_I\in \mathbb{R}^{d\times d}$ and $b_I\in \mathbb{R}^d$ are weight and bias parameters of the graph network. Similar to Eq. (\ref{eq:ac}), the calculation of $a_{ij}$ is shown as Eq. (\ref{eq:ac2}), where $v_i\in \mathbb{R}^{2d}$ is the weight vector for items.
	\begin{equation}
		a_{ij} = \cfrac{exp(\sigma(v_i[\mathbf{W}_Ih_i||\mathbf{W}_Ih_j]))}{\sum_{p\in N^{\phi_k}_i} exp(\sigma(v_i[\mathbf{W}_Ih_i||\mathbf{W}_Ih_p]))}.
		\label{eq:ac2}
	\end{equation}
	
	In addition, an order-level normalization is applied to items to obtain the graph-based representation:
	\begin{equation}
		h^\phi_i = \cfrac{1}{l}\sum_{k\leq l} c_kh^{\phi_k}_i,
		\label{eq:ol}
	\end{equation}
	where $h^\phi_i$ denotes the meta-path based representation of item $i$, $l$ is the number of meta-paths, $\mathbf{c}=\{c_1,...,c_k\}$ are weight parameters. We replace $h^\phi_i$ by $h_i$ in the rest of the paper for convenience. 
	


	\subsubsection{Mutually Exciting Temporal Network.} To capture temporal features and the fine-grained influence of social relations, STEN imports two neural temporal point processes: a mutually exciting process to model the influence of friends' behaviors from each event and a self-exciting process to model user's dynamic interests. Different from the RNN-based neural point process model, STEN utilizes the attention mechanism, which could avoid the disappearance of long-distance dependence.
	
	Specifically, STEN divides the mission into two steps: 1) Model the mutual influence of other social-connected users with the target user through mutually exciting temporal network; 2) Extract the user temporal feature from her own sequence through self-exciting temporal network. 
	
	Firstly, we model the target user $n$ and friends' event sequences as temporal point processes, denote as  $\mathcal{P}_{n}(\lambda(t|H_t))$,  $\{\mathcal{P}_{j}(\lambda(t|H_t))\}_{j=1}^M$, where $M$ is the maximum number of friends. $|\mathcal{P}_j|$ denotes the length of event sequence for user $j$. In each temporal point process, an event $e_t^j$ means that user $j$ has an interactive behavior with item at time $t$. To measure friends' influence for each event timestamp in target behavior sequence, STEN models a mutually exciting process for the target user, defined as:
	
	\begin{equation}
		\lambda^{\mathcal{m}}_n(t_i|H_{t_i}^n) = \alpha \sum_{j=1}^M\int_0^{t_i}\lambda_{n,j}(t_i,t|H_{t_i}^n,H_t^j)dt,
		\label{eq:mean}
	\end{equation}
	where $H_{t}^j$ is the condition of user $j$'s behavior history before time $t$,  $\lambda_{n,j}(t_i,t)$ represents the mutual influence intensity of friend $j$ at time $t$ on the target user at time $t_i$. $\alpha$ is a normalized parameter. For users whose friends exceed $M$, we adopt a simple sampling strategy to randomly select $M$ friends. Because Eq. (\ref{eq:mean}) explores the impact on the target user, we assume that the friends' events are independent of each other. Based on that assumption, the mutually exciting temporal network could treat each event separately. The mutual influence intensity of each friend can be regarded as a sum of the influence of different events and the mutual influence conditional intensity with two events $e_{t_i}^n, e_{t_w}^j$ follows the form of:
	
	\begin{equation}
		\lambda^{\mathcal{m}}_n(t_i|H_{t_i}^n) = \alpha \sum_{j=1}^M\sum_{w=1}^{|\mathcal{P}_j|} \Phi(e_{t_i}^n, e_{t_w}^j),
		\label{eq:matt}
	\end{equation}
	where $\Phi(e_{t_i}^n, e_{t_w}^j)$ is the cumulative mutual influence intensity function for events $e_{t_i}^n, e_{t_w}^j$ and defined as follows:  
	\begin{equation}
		\Phi(e_{t_i}^n, e_{t_w}^j) = \int_{t_w}^{t_i}\lambda_{n,j,w}(t_i,t|H_{t_i}^n,H_t^j)dt.
		\label{eq:mint}
	\end{equation}
	
	Inspired by\cite{omi2019fully}, we propose to model the cumulative mutual influence intensity function $\Phi(e_{t_i}^n, e_{t_w}^j)$, rather than the mutual influence intensity. Compared with the latter, our approach has less limitation in flexible modeling by avoiding costly approximate calculation of integral of mutual influence condition intensity shown in Eq. (\ref{eq:mean}). To take both the content of events and time interval into consideration, the cumulative mutual influence intensity function contains three components: the correlation between two events, the temporal relationship and base intensity, which is defined as follows:
	
	\begin{equation}
		\Phi(e_{t_i}^n, e_{t_w}^j)=\mathcal{F}(   \underbrace{corr(e_{t_i}^n, e_{t_w}^j)}_{correlation} + \underbrace{\beta(t_i-t_w)}_{temporal}+\underbrace{\mu}_{base}), 
		\label{eq:mi}
	\end{equation}
	where $\mathcal{F(\cdot)}$ is a mapping function: $\mathbb{R} \rightarrow \mathbb{R}^+$ to guarantee to obtain the positive cumulative mutual influence intensity, and we use softplus here. To consider the influence of friends' events, we concatenate friends' event embedding together as $\mathbf{S_F}\in\mathbb{R}^{d_F\times d}, d_F=M\times l_m$ and $l_m$ denotes the maximum length of the event sequence. Based on Eq. (\ref{eq:matt}) to (\ref{eq:mi}), we have:
	\begin{equation}
		A^	\mathcal{m} = Softmax(mask(\cfrac{\mathbf{Q}_\mathcal{m}\mathbf{K}_\mathcal{m}^T}{\sqrt{d}}+\beta_\mathcal{m}\Delta \mathbf{t}+\mu_\mathcal{m} ))),\\
		\label{eq:matt2}
	\end{equation}
	
	\begin{equation}
		\lambda^{\mathcal{m}}_n(t_i|H_{t_i}^n) = \sum_{k = 1}^{d_F}\alpha^\mathcal{m}_k\mathbf{W}_\mathcal{m}^Vh^F_k,
	\end{equation}
	where $\mathbf{Q}_\mathcal{m}=h_i^n\mathbf{W}_\mathcal{m}^Q$, $\mathbf{K}_\mathcal{m}= \mathbf{S}_F\mathbf{W}_\mathcal{m}^K$ are query and key matrices transformed by $h_i^n$ and $\mathbf{S}_F$.  $\mathbf{W}_\mathcal{m}^Q,\mathbf{W}_\mathcal{m}^K,\mathbf{W}_\mathcal{m}^V\in\mathbb{R}^{d\times d}$, $\beta_\mathcal{m}, \mu_\mathcal{m}\in \mathbb{R} $ are trainable parameters of networks. $\alpha^\mathcal{m}_k\in \mathbf{A}^\mathcal{m}$ is the mutual attention score for friends' events. $h_i^n\in \mathbf{s_n}$, $h_k^F\in \mathbf{S_F}$ are the representations of interactive item of target user and her friends learned from social heterogeneous graph, respectively. $\Delta \mathbf{t}$ denotes the time interval between the friends' events and the target event $e^n_{t_i}$. Besides, we apply a mask operation on the temporal attention network to avoid leveraging the ``future information''. In details, for target event $e_{t_i}^n$, we mask all friends' events that happened after time $t_i$.
	
	After the process of mutually exciting temporal network, STEN has extracted the influence of friends on the target user's point process, denotes as  $\mathcal{P}_{n}(\lambda_n^\mathcal{m}(t|H_t))= T_n^\mathcal{m}=\{\lambda_{n,i}^\mathcal{m}\}_{i=1}^{|\mathcal{P}_n|}$, $\lambda_{n,i}^\mathcal{m}$ is the social mutual temporal representation of the $i$-th  event for user $n$. However, there still leaves efforts on modeling the target user's event sequence to obtain her temporal representation.

	\subsubsection{Self-Exciting Temporal Network.} The goal of a self-exciting temporal network is to fuse the sequence of temporal representations for interactive items to generate the dynamic representation of users. The main idea of this structure follows the self-exciting process, which is combined with the attention mechanism. For the target process $\mathcal{P}_n(\lambda_n(t_i|H_{t_i}^n))$, we apply the attention mechanism to capture the dependency from the sequence without the limitation of length:
	\begin{equation} 
		\label{eq:sen}
		T_n^{\mathcal{S}} = Softmax(mask(\cfrac{\mathbf{Q}_\mathcal{S}\mathbf{K}_\mathcal{S}^T}{\sqrt{d}}+\beta_\mathcal{S} \Delta \mathbf{t}+\mu_\mathcal{S}))\mathbf{V}_\mathcal{S},
	\end{equation} 
	where $\mathbf{Q}_\mathcal{S}=T^\mathcal{m}_n\mathbf{W}_\mathcal{S}^Q, \mathbf{K}_\mathcal{S}=T^\mathcal{m}_n\mathbf{W}_\mathcal{S}^K$,  $\mathbf{V}_\mathcal{S}=T^\mathcal{m}_n\mathbf{W}_\mathcal{S}^V$ are query, key and value matrices obtained by linear transformations of $T^\mathcal{m}_n$. $\mathbf{W}_\mathcal{S}^Q$, $ \mathbf{W}_\mathcal{S}^K$, $ \mathbf{W}_\mathcal{S}^V\in \mathbb{R}^{d\times d}$, $\beta_\mathcal{S}, \mu_\mathcal{S}\in \mathbb{R}$ are trainable parameters. Similar to the mutually exciting temporal network, we also apply temporal mask operation on the sequence to avoid leveraging ``future information''. From Eq. (\ref{eq:sen}), we obtain the discrete value of condition intensity function over events $T_n^\mathcal{S}=\{\lambda_{n,i}^\mathcal{S}\}_{i=1}^{|\mathcal{P}_n|}$. We use the output of the last term as the temporal representation of user $h_s$.
	\subsubsection{Model Prediction}
	In addition, to model general social demands, we utilize the target user embedding $h_n$ learned from social heterogeneous graph as the query and apply a vanilla attention mechanism and average pooling to the target user interacted item sequence $I^n=\{h^n_1,h^n_2,...h_l^n\}$, which is also generated by social heterogeneous graph, shown as:
	\begin{equation}
		\label{eq:va}
		h_g = \sum Softmax(\cfrac{\mathbf{Q}_g\mathbf{K}_g^T}{\sqrt{d}})\mathbf{V}_g,
	\end{equation}
	where $\mathbf{Q}_g=h_n\mathbf{W}_g^Q, \mathbf{K}_g=I^n\mathbf{W}_g^K,\mathbf{V}_g=I^n\mathbf{W}_g^V$ are query, key and value matrices transformed by $h_n, I^n$. $\mathbf{W}_g^Q, \mathbf{W}_g^K, \mathbf{W}_g^V\in \mathbb{R}^{d\times d}$ are trainable parameters in transformation. 
	The final representation of user is obtained by a linear combination of the two parts: temporal behaviors and social influence, shown as follows:
	\begin{equation}
		h_{hybrid}=\mathbf{W}_u(h_s||h_g),
		\label{eq:up}
	\end{equation}
	where $||$ denotes concatenation operation, $\mathbf{W}_u\in \mathbb{R}^{d\times 2d}$ is a transformation matrix and $h_{hybrid}$ is the final representation of the user. The prediction of next item is obtained through a softmax function. 
	\begin{equation}
		Y = Softmax(h_{hybrid}\mathbf{I}^T),
	\end{equation}
	where $\mathbf{I}\in \mathbb{R}^{|I|\times d}$ is the embedding matrix of items, $Y$ denotes the probabilities of items will be selected as the next one. 
	
	\subsection{Model Optimization}
	In the training stage, we adopt the Bayesian Personalized Ranking (BPR) framework\cite{rendle2012bpr} for optimizing the recommendation task:
	\begin{equation}
		\label{eq:lossbpr}
		\mathcal{L}_{BPR} = \sum_{\hat{y}_{ui^+}\in D_{t},\hat{y}_{ui^-}\in D_{n}}-ln\sigma(\hat{y}_{ui^+}-\hat{y}_{ui^-}),
	\end{equation}
	where $D_t, D_n$ denote the training set and unobserved item set, $\hat{y}_{ui^+}$ is the predicted probability of observed item $i^+$ and $\hat{y}_{ui^-}$ is the predicted probability of the sampled unobserved item $i^-$. To effectively learn from the temporal sequence, we consider the fitting of temporal point process by optimizing the maximum log-likelihood of observed items:
	\begin{equation}
		\label{eq:losstpp}
		\begin{aligned}
			\mathcal{L}_{\mathcal{P}_{n}} &= -\sum_{i = 1}^{n} log(\lambda(t_i|H_{t_i})) +\int_{0}^{T}\lambda(t|H_t)dt \\
			&= -\sum_{i = 1}^{n} \left[ log(\lambda(t_i|H_{t_i})) +\int_{t_i}^{t_{i+1}}\lambda(t|H_t)dt \right]
			\\
			&= -\sum_{i = 1}^{n} log(\lambda_{n,i}^\mathcal{S})+ \frac{1}{2}\sum_{i = 1}^{n} \Delta t_i(\lambda_{n,i}^\mathcal{S}+\lambda_{n,i+1}^\mathcal{S}), 
		\end{aligned}	
	\end{equation}
	where $\Delta t_i$ is the time interval between observed items. It leverages trapezoidal rule, one of numerical approximations, to calculate the integral in Eq. (\ref{eq:losstpp}). We finally adopt joint learning for optimizing both the loss of BPR and temporal point process, which is defined as Eq. (\ref{eq:loss}):
	\begin{equation}
		\label{eq:loss}
		\mathcal{L} = \mathcal{L}_{BPR} + \mathcal{L}_{\mathcal{P}_{n}} + \gamma||\Theta||_2,
	\end{equation}
	where $\gamma$ and $\Theta$ are regularization weights and the set of trainable parameters, respectively. 
	
	\section{Experiments}
	%
	%
	
	\subsection{Experimental Setup}
	
	\subsubsection{Datasets.} To evaluate the performance of our proposed model and other baselines, we conduct experiments on three real-world datasets. We provide descriptive statistics of them in Table \ref{tab:dataset}.
	\begin{table}[h]
		\setlength{\abovecaptionskip}{-5pt} 
		\setlength{\belowcaptionskip}{-5pt} 
		\caption{Basic Statistic of Datasets}
		\vspace{0.3cm}
		\centering
		\renewcommand{\arraystretch}{1.2}
		\label{tab:dataset}
		\setlength{\tabcolsep}{1mm}{
			\resizebox{0.4\textwidth}{!}{
				\begin{tabular}{ c c c c }
					\toprule[1pt]
					Dataset  &Delicious  &Yelp &Ciao\\
					\hline
					\#users & 1,867 & 74,982 &6,596 \\
					\#items & 24,052 & 59,692 &107,320\\
					\#Events & 437,593 & 811,503 & 274,065 \\
					\#Social Relations & 15,328&2,577,679 &114,893\\
					Avg. events per user & 234.38 &10.82 & 41.55\\
					Avg. relations per user & 8.24 & 34.37 & 17.42\\
					\bottomrule[1pt]
			\end{tabular}}
		}
	\end{table}
	
	\textit{Delicious.}\footnote{https://grouplens.org/datasets/hetrec-2011/} This dataset contains social networking and bookmarking information from Delicious online social bookmarking system. It involves the sequence of tagging for each user with timestamps and relationships between users.
	
	\textit{Yelp.}\footnote{https://www.yelp.com/dataset} This dataset contains abundant users' reviews for local businesses (it is also timestamped) on yelp platform. Here we treat each review as an interactive behavior with a business. 
	
	\textit{Ciao.}\footnote{https://www.cse.msu.edu/~tangjili/datasetcode/truststudy.htm} This dataset involves users' rating records and social relations, \textit{e.g.}, trust relations with other users.
	
	For data preprocessing, we follow the common previous works\cite{sun2019bert4rec,kang2018self} to convert all numeric ratings and reviews to implicit feedback with value 1. To avoid too large period, we choose first 10\% data as a division for Yelp dataset to do experiments. To guarantee the quality of datasets, we filter out users that interacted less than 5 times and items appear less than 10 times for Yelp and 5 for the others. For all datasets, we use the last item of each sequence for testing and use the penultimate item for validation.
		
	\begin{table*}[t]
		\caption{Comparison of STEN and baselines over three datasets. The boldface indicates the highest score and the underline indicates the second highest score. BPR-MF is conventional baseline. FPMC, SRGNN, SASRec and Bert4Rec are general sequential baselines. THP and TiSASRec are time-aware sequential baselines. SBPR, DANSER and Diffnet are social baselines.}
		\setlength{\abovecaptionskip}{-5pt} 
		\setlength{\belowcaptionskip}{-5pt} 
		\label{tab:rq1}
		\vspace{0.15cm}
		\centering
		\begin{threeparttable}
			\renewcommand{\arraystretch}{1.1}
			\setlength{\tabcolsep}{2mm}{
				\resizebox{\textwidth}{!}{
					\begin{tabular}{cccccccccccccc}
						\toprule[1pt]
						\multirow{2}{*}{Dataset}&\multirow{2}{*}{Metrics} &Conventional & \multicolumn{4}{c}{Sequential}& \multicolumn{2}{c}{Time-aware} &\multicolumn{3}{c}{Social} &\multirow{2}{*}{STEN}  &\multirow{2}{*}{Impr.} \\
						\cmidrule(r){3-3} \cmidrule(r){4-7} \cmidrule(r){8-9} \cmidrule(r){10-12}
						\multicolumn{1}{c}{}&\multicolumn{1}{c }{} &BPR-MF &FPMC &SRGNN &SASRec &Bert4Rec &THP &TiSASRec  &SBPR &DANSER &Diffnet \\
						\hline
						\multirow{6}{*}{Delicious} &Recall@10  
						&0.1935 	&0.2784 	&0.2370 	&0.2857 	&0.3125 	&0.2176 	&\underline{0.3362} 	&0.2338 	&0.2969 	&0.3016 	&\textbf{0.3492*} 	&+3.87\% \\
				
						\multicolumn{1}{c }{}  &Recall@20 &0.2454 	&0.3188 	&0.2703 	&0.3133 	&0.3496 	&0.2676 	&\underline{0.3695} 	&0.2661 	&0.3144 	&0.3238 	&\textbf{0.3731*} 	&+0.98\% \\
				
						\multicolumn{1}{c }{}  &NDCG@10 &0.1343 	&0.2317 	&0.1640 	&0.2145 	&0.2583 	&0.1454 	&\underline{0.2702} 	&0.1687 	&0.1809 	&0.1963 	&\textbf{0.2833*} 	&+4.83\%\\
						\multicolumn{1}{c }{}  &NDCG@20 &0.1475 	&0.2419 	&0.1724 	&0.2215 	&0.2653 	&0.1562 	&\underline{0.2785} 	&0.1792 	&0.1849 	&0.2031 	&\textbf{0.2885*} 	&+3.58\%\\ 
						\multicolumn{1}{c }{} &MRR@10 &0.1161 	&0.2170 	&0.1411 	&0.1918 	&0.2353 	&0.1287 	&\underline{0.2528} 	&0.1418 	&0.1535 	&0.1563 	&\textbf{0.2673*} 	&+5.74\%\\ 
						\multicolumn{1}{c }{} &MRR@20 &0.1197 	&0.2197 	&0.1434 	&0.1937 	&0.2385 	&0.1291 	&\underline{0.2547} 	&0.1446 	&0.1568 	&0.1646 	&\textbf{0.2692*} 	&+5.70\%\\ 
						\hline
						\multirow{6}{*}{Yelp} &Recall@10 &0.0353 	&0.0436 	&0.0367 	&0.0461 	&0.0452 	&0.0395 	&\underline{0.0493} 	&0.0375 	&0.0451 	&0.0453 	&\textbf{0.0512*} 	&+3.80\%\\
						\multicolumn{1}{c }{}  &Recall@20	&0.0589 	&0.0739 	&0.0620 	&0.0758 	&0.0756 	&0.0649 	&\underline{0.0815} 	&0.0624 	&0.0746 	&0.0763 	&\textbf{0.0846*} 	&+3.81 \%\\
						\multicolumn{1}{c }{}  &NDCG@10	&0.0177 	&0.0212 	&0.0183 	&0.0233 	&0.0226 	&0.0191 	&\underline{0.0247} 	&0.0185 	&0.0227 	&0.0226 	&\textbf{0.0262*} 	&+6.29 \%\\
						\multicolumn{1}{c }{}  &NDCG@20	&0.0236 	&0.0288 	&0.0246 	&0.0308 	&0.0302 	&0.0255 	&\underline{0.0327} 	&0.0247 	&0.0301 	&0.0303 	&\textbf{0.0344*} 	&+5.09 \%\\
						\multicolumn{1}{c }{}  &MRR@10	&0.0124 	&0.0145 	&0.0128 	&0.0165 	&0.0158 	&0.0131 	&\underline{0.0172} 	&0.0128 	&0.0160 	&0.0161 	&\textbf{0.0185*} 	&+7.32 \%\\
						\multicolumn{1}{c }{}  &MRR@20	&0.0140 	&0.0166 	&0.0145 	&0.0185 	&0.0178 	&0.0148 	&\underline{0.0194} 	&0.0145 	&0.0180 	&0.0182 	&\textbf{0.0207*} 	&+6.54 \%\\
						 \hline
						\multirow{6}{*}{Ciao} &Recall@10&0.0591 	&0.0593 	&0.0552 	&0.0624 	&0.0832 	&0.0676 	&\underline{0.0960} 	&0.0705 	&0.0815 	&0.0897 	&\textbf{0.1019*} 	&+5.79 \%\\ 
						\multicolumn{1}{c }{}  &Recall@20 &0.0829 	&0.0912 	&0.0871 	&0.0972 	&0.1048 	&0.0976 	&0.1220 	&0.0984 	&0.1035 	&\underline{0.1232} 	&\textbf{0.1294*} 	&+4.79 \%\\ 
						\multicolumn{1}{c }{}  &NDCG@10&0.0304 	&0.0309 	&0.0301 	&0.0325 	&0.0523 	&0.0385 	&\underline{0.0564} 	&0.0369 	&0.0455 	&0.0508 	&\textbf{0.0596*} 	&+5.37 \%\\ 
						\multicolumn{1}{c }{}  &NDCG@20&0.0365 	&0.0388 	&0.0381 	&0.0413 	&0.0575 	&0.0421 	&\underline{0.0645} 	&0.0439 	&0.0515 	&0.0597 	&\textbf{0.0685*} 	&+5.84 \%\\ 
						\multicolumn{1}{c }{}  &MRR@10&0.0217 	&0.0223 	&0.0225 	&0.0235 	&0.0384 	&0.0249 	&\underline{0.0427} 	&0.0266 	&0.0344 	&0.0389 	&\textbf{0.0462*} 	&+7.58 \%\\ 
						\multicolumn{1}{c }{}  &MRR@20&0.0234 	&0.0244 	&0.0247 	&0.0259 	&0.0406 	&0.0149 	&\underline{0.0438} &0.0286 	&0.0366 	&0.0418 	&\textbf{0.0479*} 	&+8.56 \%\\ 
						\bottomrule[1pt]
				\end{tabular}}
			}
			\begin{tablenotes}
				\footnotesize
				\item \textbf{*} denotes the statistically significant improvements (\textit{i.e.}, one-tailed t-test with \textit{p} $\leq$ 0.05) compared with the best baseline.
			\end{tablenotes}
		\end{threeparttable}
			\vspace{-0.2cm}
	\end{table*}

	\subsubsection{Evaluation Metrics.}  We adopt several common evaluation metrics to evaluate the recommendation performance, including Recall, Normalized Discount Cumulative Gain (NDCG) and Mean Reciprocal Rank (MRR):
	\begin{itemize}[leftmargin=*]
		\item \textbf{Recall@k.} It means the percentage of groundtruth relevant items appear within the top-$k$ ranking list. Consider to next-item recommendation, it is equivalent to the hit ratio.
		\item \textbf{NDCG@k.} It is a standard ranking metric and reflects both the correlation and position for each recommended item. For next-item recommendation, it is formulated as NDCG $ =\frac{1}{log_2(1+r_p)}$, where $r_p$ is the rank of the positive item.
		\item \textbf{MRR@k.} It takes the actual rank of target item into consideration and is calculated by the mean value of the reciprocal of target item's actual rank. When the actual rank is out of k, the reciprocal rank is set to zero.
	\end{itemize}

	\subsubsection{Baseline Methods.} To evaluate the performance of STEN, we choose several comparative baselines, including state-of-the-art methods from both sequential and social recommendation.
	\begin{itemize} [leftmargin=*]
		\item \textbf{BPR-MF}\cite{rendle2009bpr}: It utilizes matrix factorization and rating matrix to model latent factors of users and items with the loss of Bayesian  pairwise ranking loss.
		\item \textbf{FPMC}\cite{rendle2010factorizing}: This method captures users' preference from their sequential behaviors based on Markov Chain.
		\item \textbf{SRGNN}\cite{wu2019session}: This method treats user's sequential behaviors as graph-structural data to capture features from complex transitions of items.
		\item \textbf{SASRec}\cite{kang2018self}: This method adopts an attention mechanism to capture users' sequential behaviors and makes prediction based on relatively few actions.
		\item \textbf{BERT4Rec}\cite{sun2019bert4rec}: This method employs a deep bidirectional self-attention to model user behavior sequences and achieves strong performance.
		\item \textbf{THP}\cite{zuo2020transformer}: This method is a temporal point process model, which utilizes a transformer to capture the long-term dependency in the temporal sequence. 
		\item \textbf{TiSASRec}\cite{li2020time}:TiSASRec is a time-sensitive method, which models both the absolute positions of items and the time intervals between them together. 
		\item \textbf{SBPR}\cite{zhao2014leveraging}: This method leverages social connections to enhance the performance of Bayesian Personalized Ranking. 
		\item \textbf{DANSER}\cite{wu2019dual}: This method collaboratively learns representation of user-specific attention and context-aware attention in social information by dual graph attention networks. 
		\item \textbf{Diffnet}\cite{wu2019neural}: Diffnet is a deep influence propagation model that stimulates how users are influenced by the recursive social diffusion process for social recommendation.
	\end{itemize}

	\subsubsection{Implementation Details \& Parameter Settings}
	In our experiments, we implement our model with Python 3.7 and Pytorch 1.7.0 and train it and other baselines on 4 Nvidia GTX 2080 GPU with 12G memory. In parameter settings,  we use Adam\cite{kingma2014adam} as the optimizer for all models. In addition, we also adopt mini-batch with batchsize of 256. The embedding size and learning rate for all methods are determined by grid search from \{40, 60, 80, 100, 120, 140\} and \{0.001, 0.005, 0.01 \}, respectively. The parameters are randomly initialized to fit normal distribution $\mathit{N}(0,0.01)$. To avoid overfitting and improve robustness, we add the dropout operation with dropout ratio tuning from $\{0.1, 0.2, ..., 0.9\}$ and regularization term with coefficient tuning from $\{0.0001, 0.0005, 0.001, 0.005\}$. For a fair comparison, the results for all baselines are reported with the optimal hyper-parameter settings. For Diffnet, we remove the fusion component for both user and item to accommodate the absence of features. For THP, we transfer the task of event type prediction to item recommendation and remove the time prediction loss to improve its performance.
	For our model STEN, the max-order of the graph is set to 2. The sampling number of friends is set to 10 and the maximum sequence length depends on characteristics of datasets, we set \{300, 30, 50\} for three datasets, respectively.

	\subsection{Overall Performance}
	Table \ref{tab:rq1} shows the performance of our model and baselines over three datasets under the metrics of Recall@k,  NDCG@k, MRR@k, k = \{10, 20\}. According to table \ref{tab:rq1}, we can get the following observation.

	The conventional recommendation method only takes the rating or interaction into consideration, without both the social relationship and the temporal information, which could explain why BPR-MF achieves poor performance. 

	For general sequential baselines, FPMC performs better than BPR-MF, which indicates the local sequential information is beneficial to sequential recommendation.
We believe that the reason SR-GNN performs worst in the general sequential method is that it treats each session individually. SASRec and Bert4Rec both utilize attention mechanisms to capture sequential information. Bert4Rec overcomes SASRec on two datasets, which indicates the advantage of bi-direction networks.
In addition, for time-aware sequential methods, \textit{e.g.}, THP utilizes the transformer-based neural Hawkes process to model the asynchronous temporal sequences. However, its performance is worse than other methods. It shows the limitation that it aims to obtain the event representation via the neural Hawkes process. Besides, we think the way to transfer THP into the recommendation task is too direct to be effective. 
TiSASRec extends SASRec by taking the time interval into consideration and achieves the second-highest results over three datasets, which demonstrates that time interval is an important factor in recommendation. 

For social-based methods, SBPR is worse than the other two, which indicates GNN is a strong tool to model social relations. DANSER gets great improvement by utilizing the dual attention mechanism to capture the social information for both users and items. Diffnet, which uses the defusion graph structure to stimulate social influence propagation to improve the performance, has a similar performance to the former.

Finally, our proposed model STEN achieves the best performance on three datasets and all metrics. The reasons could be summarized as follows: (1) it introduces the temporal point process as the prototype to model both sequential order and time interval; (2) the designed structure mutually exciting attention network achieves the effect of directly modeling the influence of friends sequence on the target user and capture the temporal relationships between behavior events across users; (3) it utilizes a social heterogeneous graph to generate effective social node embedding.
	
	\subsection{Ablation Study}
	To justify the effectiveness of our components, we analyze the impact of utilization of social and temporal information via an ablation study over all datasets. Table \ref{tab:rq2} shows the performance of STEN and all variants. The variants and the analysis of experiment results are listed as follows:
	
		\begin{table}[t]
		\setlength{\abovecaptionskip}{-4pt} 
	
		\caption{Ablation study on three datasets. (R@10 for Recall@10, N@10 for NDCG@10)}
		\vspace{0.2cm}
		\centering
		\label{tab:rq2}
		\renewcommand{\arraystretch}{1.2}
		\
		\setlength{\tabcolsep}{1.5mm}{
			\resizebox{0.45\textwidth}{!}{
				\begin{tabular}{ccccccc}
					\toprule[1pt]
					\multirow{2}{*}{Method} &\multicolumn{2}{c}{Delicious}& \multicolumn{2}{c}{Yelp}&\multicolumn{2}{c}{Ciao} \\
					\cmidrule(r){2-3} \cmidrule(r){4-5} \cmidrule(r){6-7}	
					&R@10&N@10&R@10&N@10&R	@10&N@10\\
					\hline
					w/o GE&0.3355&0.2722&0.0468&0.0229&0.0976&0.0582\\ 
					w/o MT&0.3201&0.2646&0.0431&0.0215&0.0833&0.0490\\
					w/o ST&0.3075&0.2529&0.0447&0.0221&0.0815&0.0482\\
					w/o TC&0.3163&0.2617&0.0463&0.0232&0.0826&0.0484\\
					Default&\textbf{0.3492}&\textbf{0.2833}&\textbf{0.0512}&\textbf{0.0262}&\textbf{0.1019}	&\textbf{0.0596}\\
					\bottomrule[1pt]
			\end{tabular}}
		}
		\vspace{-0.5cm}
	\end{table}

	\textit{(a) Remove Social \underline{G}raph \underline{E}mbedding Layer.} In our model STEN, social information is utilized by two components,  the social heterogeneous graph network, and mutually exciting attention network. We replace the structural graph embedding with a trainable embedding lookup operation. The results show that it suffers a clear decrease compared to the default model, which denotes the structural view of social relationships is important to describe user and item representations. 

	\textit{(b) Remove \underline{M}utual-Excitation \underline{T}emporal Networks.}  Here we remove the part of mutually exciting and directly utilize the social graph embedding to calculate the event conditional intensity. The result shows that it suffers a significant drop. The phenomenon that it performs worse than \textit{(a)} could demonstrate the superiority of the direct paradigm. Without the mutually exciting temporal networks, the model degrades to a user-level model and suffers a lot of information loss.

	\textit{(c) Remove \underline{S}elf-Excitation \underline{T}emporal Networks.} Here we replace the self-exciting module with average pooling. We can observe that it performs the worst among all variants. The result draws to the fact that the impact of the association between user's own behaviors is more important than the impact of friend's behaviors.

	\textit{(d) Remove \underline{T}emporal \underline{C}omponent.} To test the effectiveness of leveraging temporal information, here we remove the temporal component in STEN. Specifically,  we remove the temporal term in Eq. (\ref{eq:mint}) and the temporal loss function $\mathcal{L}_{\mathcal{P}_{n}}$. The results indicate that removing the temporal component leads to significant performance degradation. We believe when temporal information is lost, the model cannot capture the specific relationship between events. 
	
	\subsection{Sensitivity Analysis}
	
		\begin{figure}[t]
		\setlength{\abovecaptionskip}{0pt} 
		\setlength{\belowcaptionskip}{-0.5cm} 
		\centering  
		\subfigure{
			\includegraphics[width=0.231\textwidth]{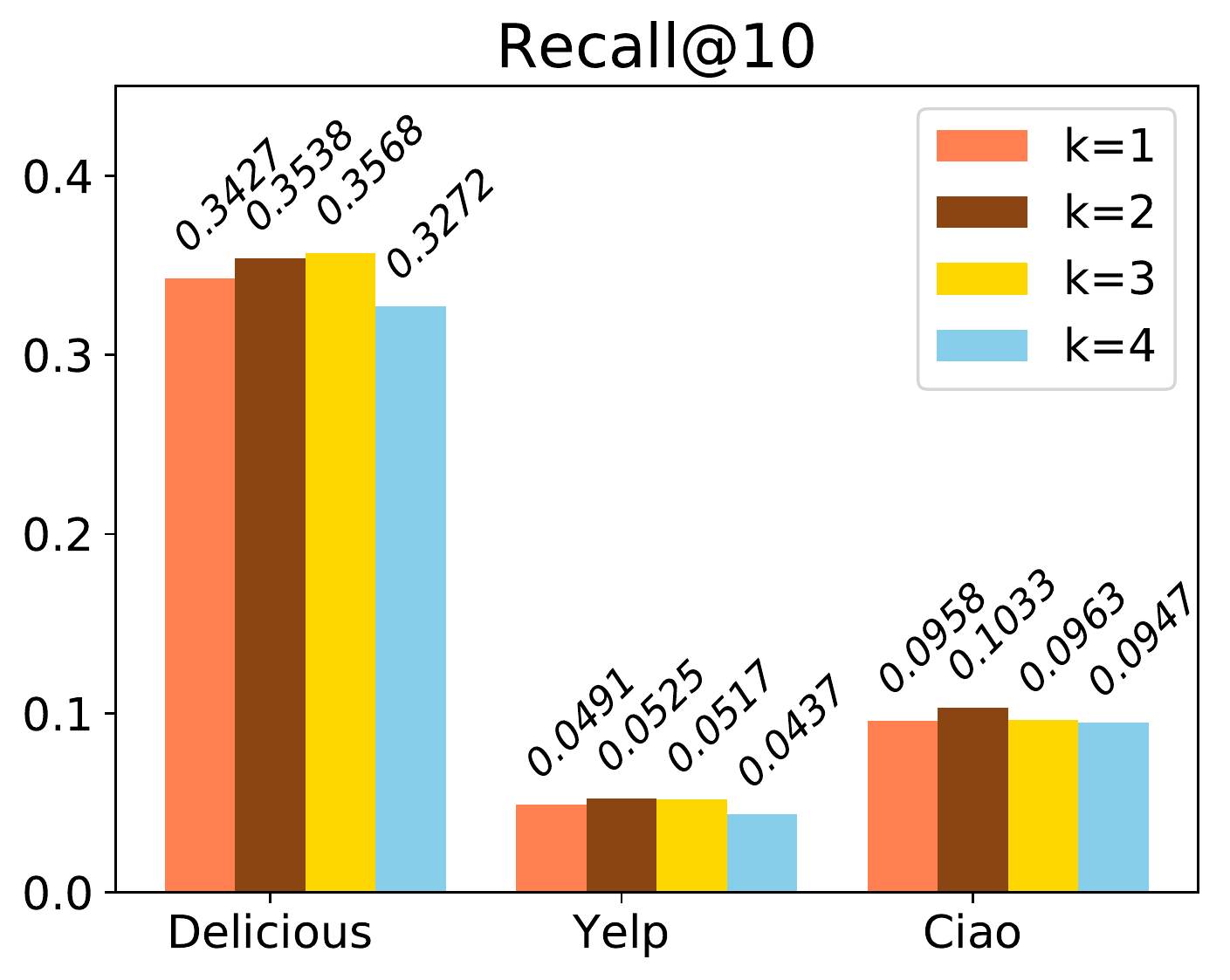}}
		\subfigure{
			\includegraphics[width=0.231\textwidth]{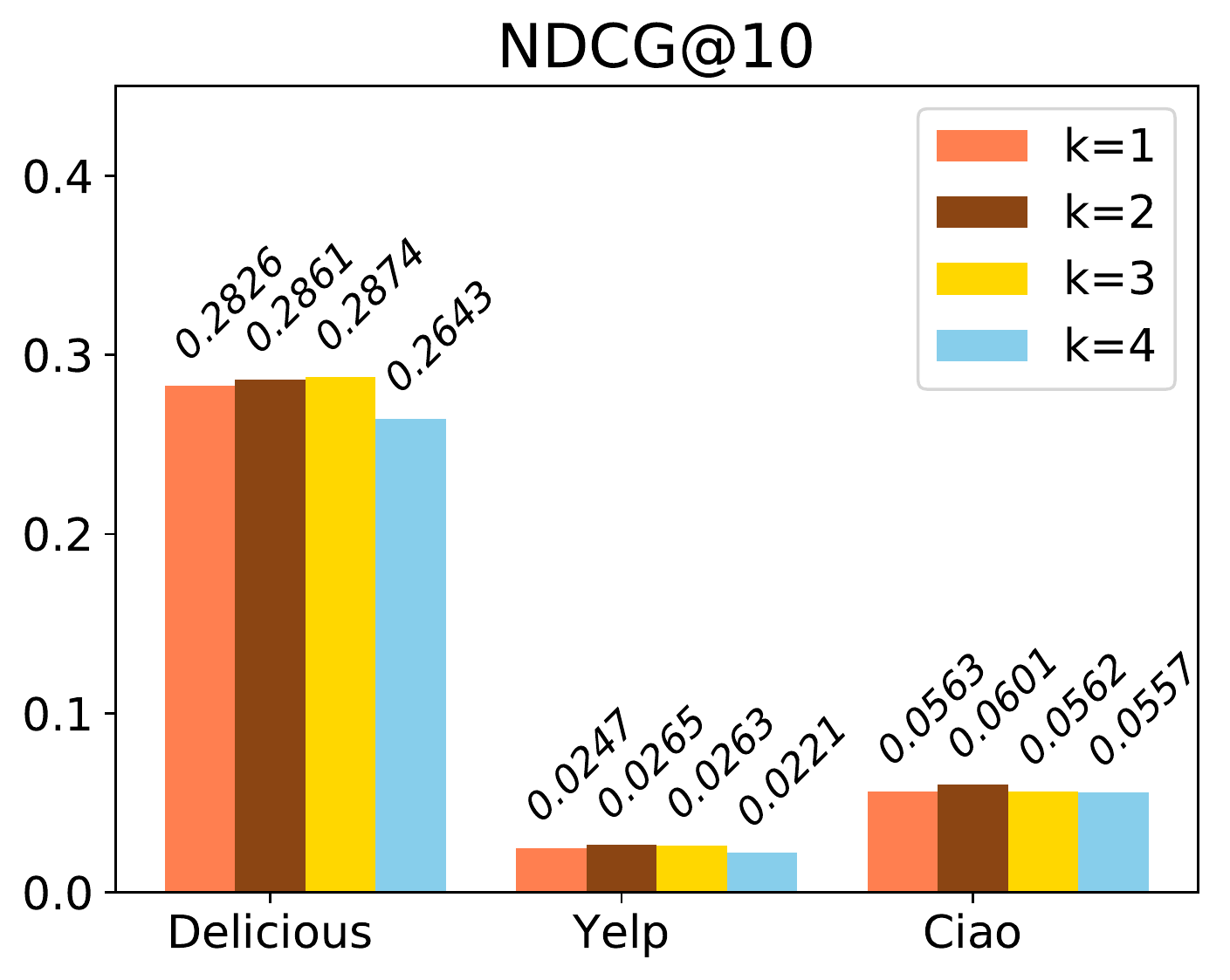}}
		\caption{The impact of the order of neighbors}
		\label{fig:on}
		\setlength{\belowcaptionskip}{-1cm} 
	\end{figure}

	\subsubsection{Performance Comparison w.r.t. the Order of Neighbors} As discussed in Section 3.1.1, we test the impact of a different order of neighbors in the social heterogeneous graph. Here we choose order k$\in\{1,2,3,4\}$ to do experiments and the higher-order social relation graph involves the information of the lower-order neighbors.  
	Figure \ref{fig:on} illustrates the performance under the condition of a different order of neighbors. We can find that the performance is improved by the second-order graph. We analyze it for the reason that the second-order graph provides extra information on social connections and contributes to social sparsity. However, when the order becomes larger, the results change to worse. We consider that it is because the higher-order graph leads to graph convergence and noise, \textit{i.e.}, the differences between users are narrowed. 
	

	\subsubsection{Performance Comparison w.r.t. the Window Size of Friend Sequence.} 
\begin{figure}[t]
		\setlength{\abovecaptionskip}{-2pt} 
		\centering  
		\subfigure{
			\includegraphics[width=0.15\textwidth]{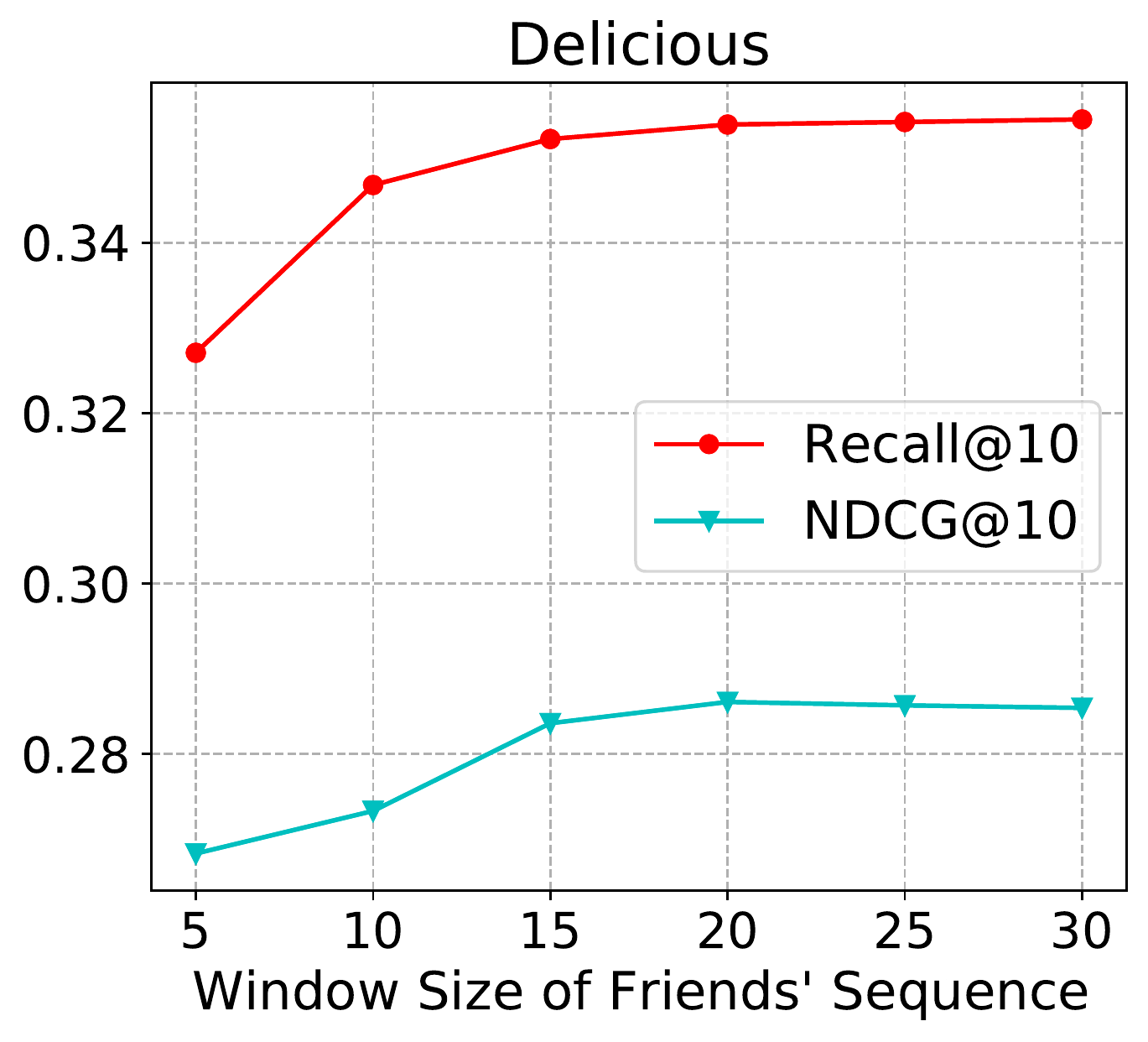}}
		\subfigure{
			\includegraphics[width=0.15\textwidth]{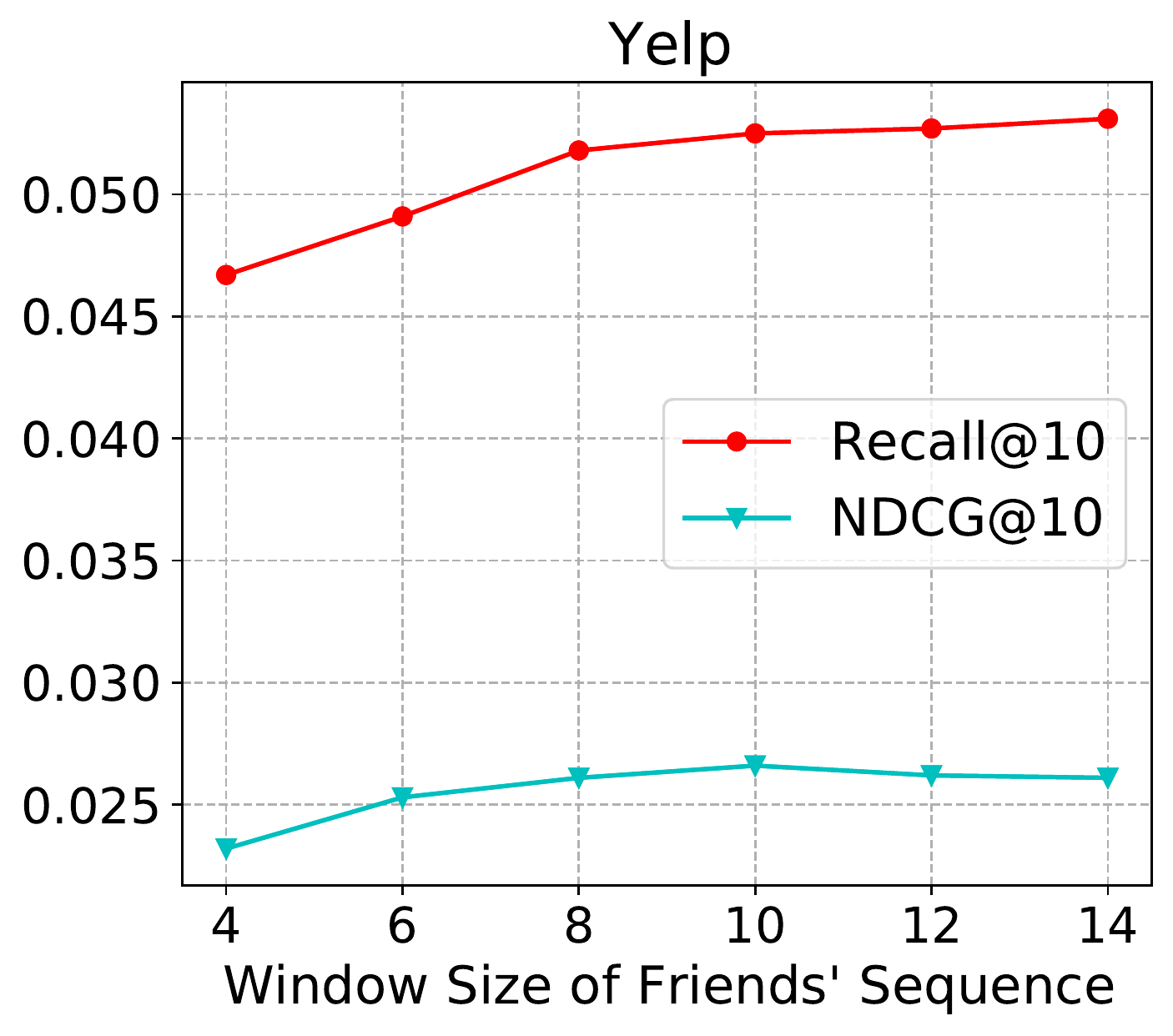}}
		\subfigure{
			\includegraphics[width=0.15\textwidth]{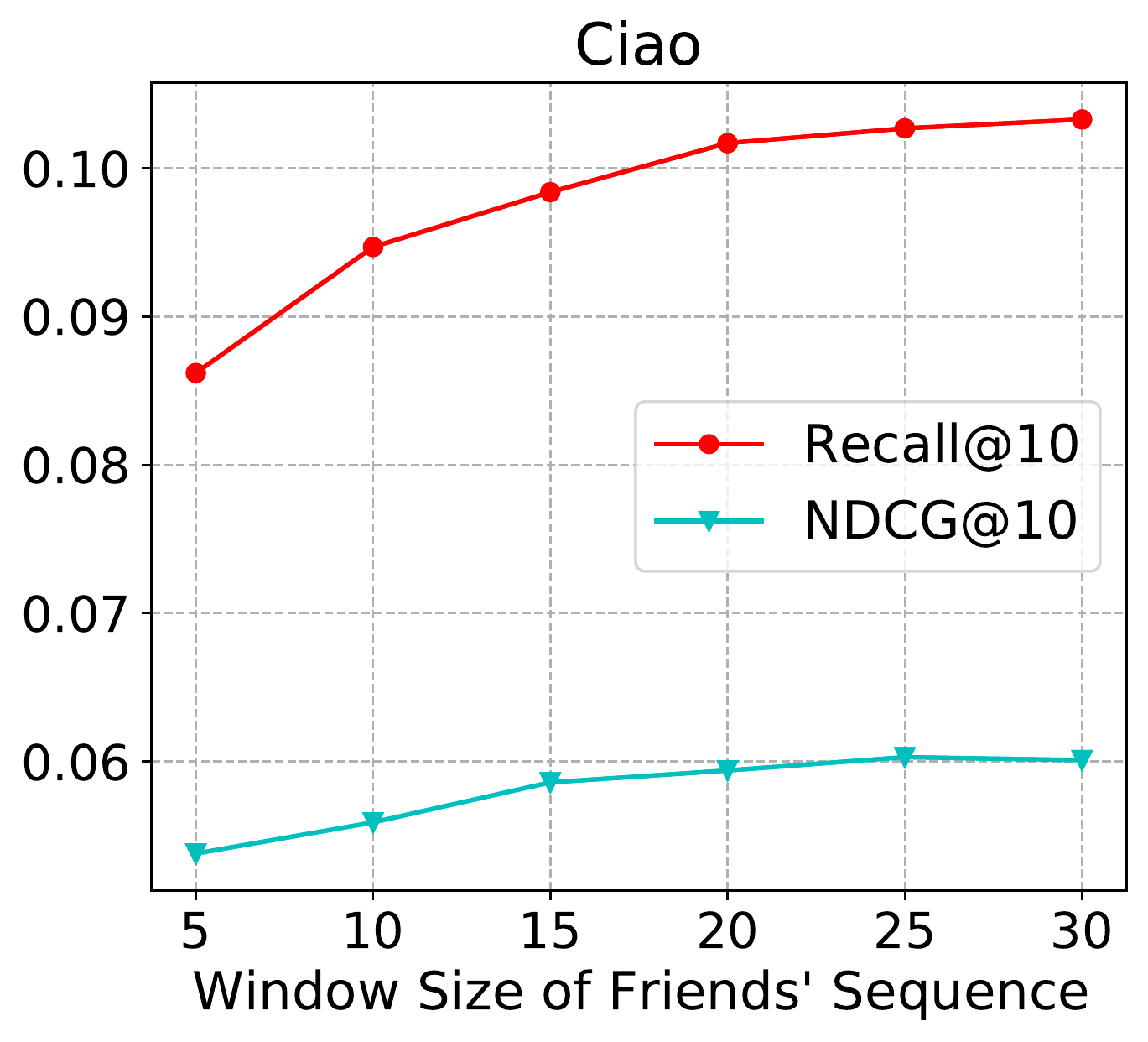}}
		\caption{The impact of the window size of friend sequences.}
		\label{fig:ws}
		\setlength{\belowcaptionskip}{-10pt}

	\end{figure}

STEN extracts the event-level social influence through the mutually exciting temporal network. The larger window size brings richer social information and more complicated calculations. Hence, we explore the impact of the window size of friend behavior event sequence. As Figure \ref{fig:ws} illustrates, with the increasing of the window size, the performance is first improved and then converges. When the window size is too short, the model only captures the current state of the friend, which contains a large deviation. The selection of window size depends on the average length of the user's event sequence.
	
	\subsubsection{Performance Comparison w.r.t. the Number of Sampled Friends} 
\begin{figure}[t]
		\setlength{\abovecaptionskip}{-2pt} 
		\centering  
		\subfigure{
			\includegraphics[width=0.15\textwidth]{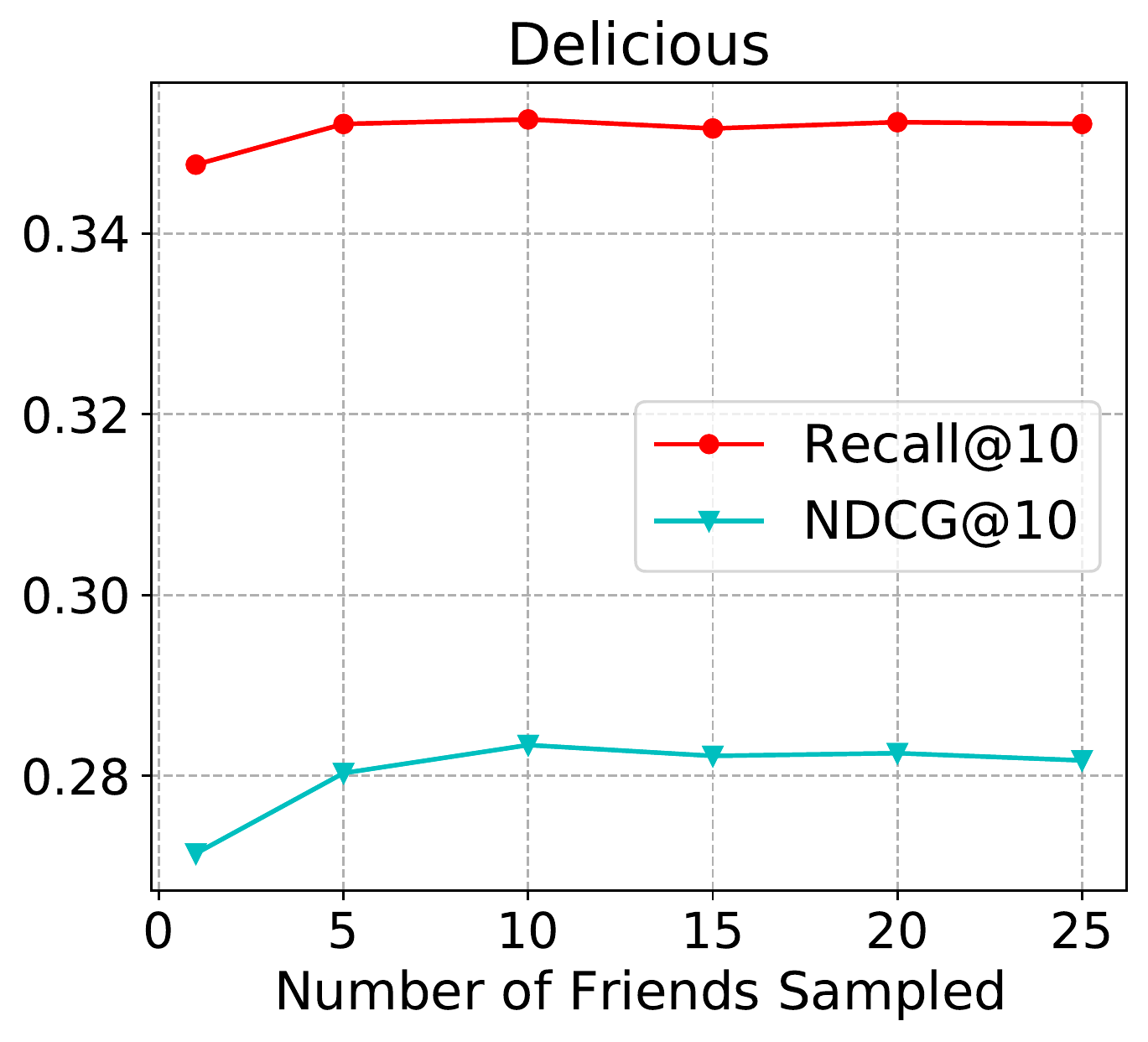}}
		\subfigure{
			\includegraphics[width=0.15\textwidth]{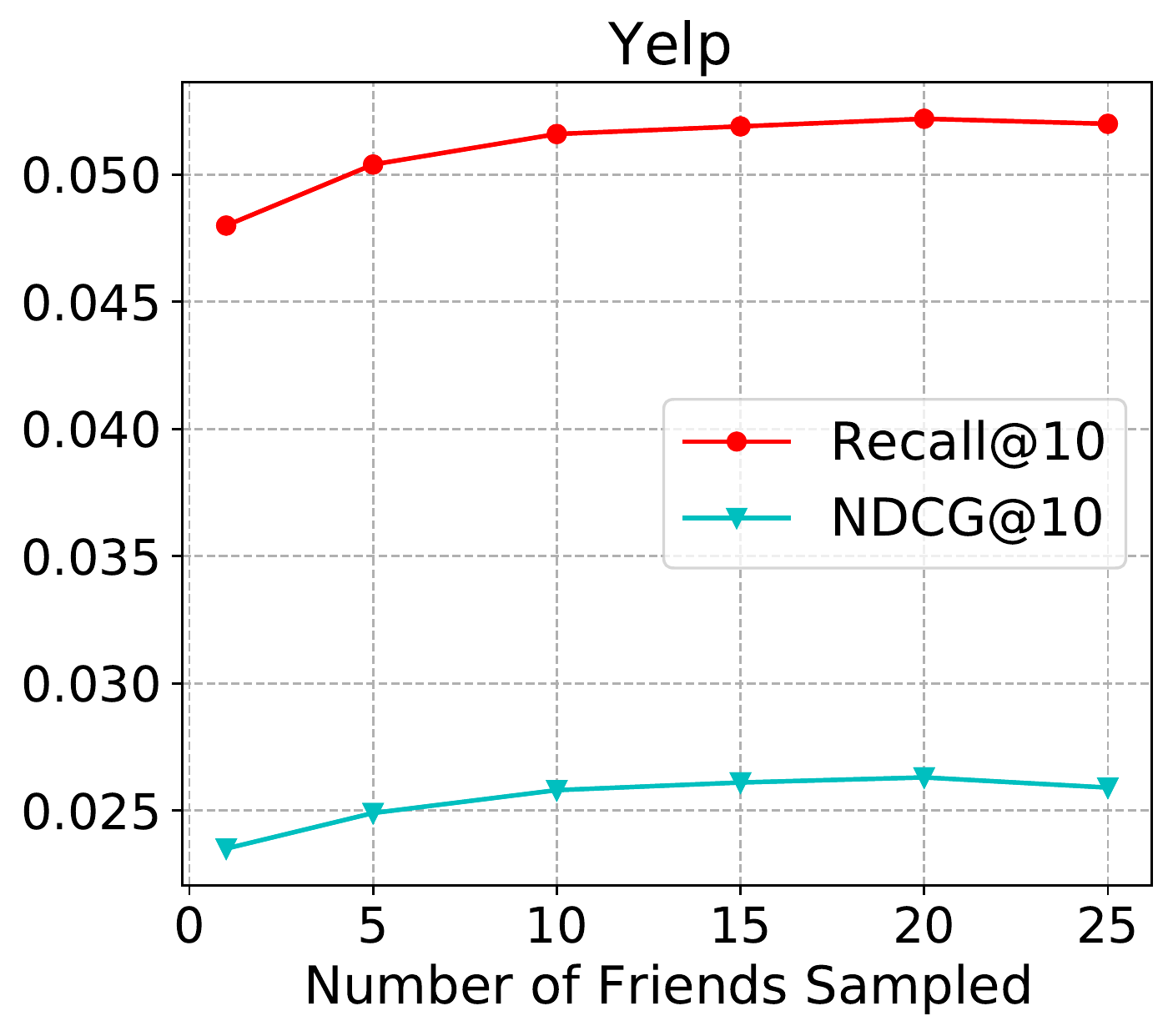}}
		\subfigure{
			\includegraphics[width=0.15\textwidth]{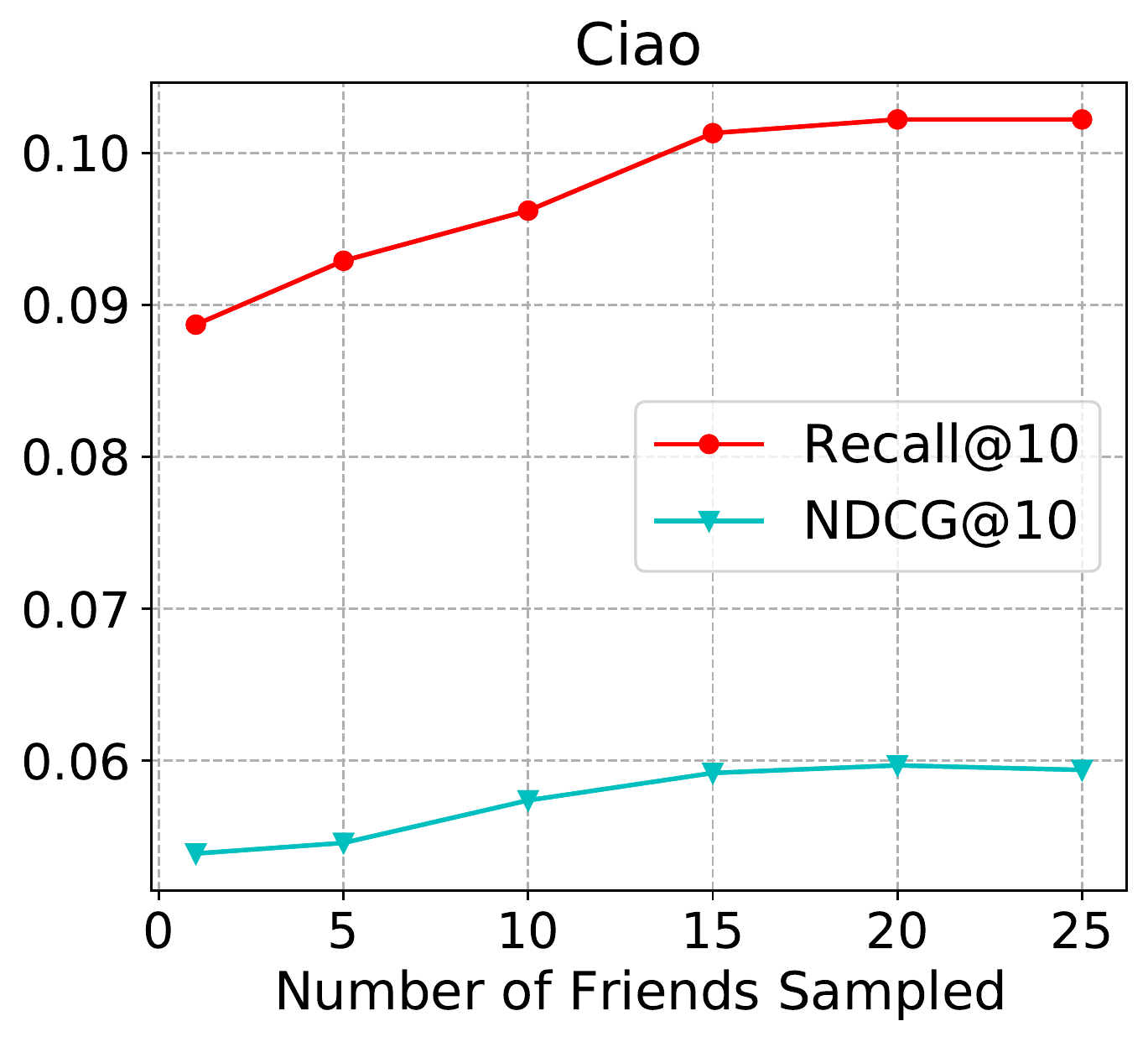}}
		\caption{The impact of the number of sampled friends.}
		\label{fig:ns}
		\setlength{\belowcaptionskip}{-2cm}

	\end{figure}

To reduce the computation for users with complex relationships, STEN adopts a random sampling operation for their friends in mutually exciting temporal network. Figure \ref{fig:ns} illustrates how the model's performance varies with the number of sampled friends. The results show that the performance of the model improves first and then converges with the increase of the number of samples. When only one friend is sampled, the performance of the model decreases significantly. Compared with the rest, the result on Delicious shows it is insensitive to the number of sampled friends. We consider the reasons can be summed up into two points: 1) As Table \ref{tab:dataset} shows, the average relations is obviously less than the rest. 2) It has a smaller proportion of Time-Efficient Friends (TEF). The definition of TEF is that the sequence of behavior events has an intersection with the target user in time. The proportions of TEF over three datasets are shown in Table \ref{tab:tef}.
	
	\begin{table}[h]
		\setlength{\abovecaptionskip}{0cm}
		\caption{The proportions of time-efficient friends}
		\vspace{0.2cm}
		\centering
		\renewcommand{\arraystretch}{1.2}
		\label{tab:tef}
		\setlength{\tabcolsep}{3mm}{
			\begin{tabular}{ c c c c}
				\toprule[1pt]
				Dataset &Delicious &Yelp &Ciao\\
				\hline
				TEF(\%) &59.16 &97.75 &94.93\\
				\bottomrule[1pt]
		\end{tabular}}
			\vspace{-0.5cm}
	\end{table}

	\subsection{Model Explainability}
\begin{figure}[t]
		\setlength{\belowcaptionskip}{-0.3cm} 
		\setlength{\abovecaptionskip}{0.1cm}
		\centering  
		\includegraphics[width=0.5\textwidth]{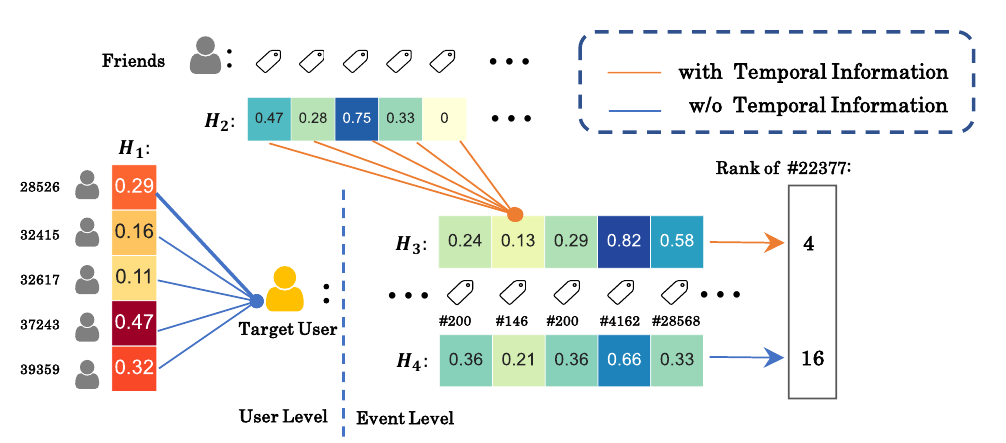}
		\caption{Visualization of attention weights in STEN. $H_1$ denotes the impact of friends to target user. $H_2$ denotes the impact of friends events on target user event. $H_3$, $H_4$ denote the impact of target user behavior events on themselves, where $H_3$ involves the temporal information, but $H_4$ does not. }
		\label{fig:cs}
	\end{figure}

	In this part, we conduct a case study on the Delicious dataset to show the explainability of our model. We visualize the attention weights of different modules in Figure \ref{fig:cs}. The value in $H_1$, which is obtained from the graph network component, indicates the impact of friends on to target user. It shows the coarse-grained relationships at the user level, while $H_2,H_3,H_4$ describe the fine-grained relationships at the event level. $H_2$ is obtained from mutually exciting temporal network, which indicates the impact of a friend's events on event \#146. The value 0 in $H_2$ means that the event happens after event \#146, which obeys the rule that no future information is leveraged. $H_3, H_4$ indicate the impact of target user behavior events on themselves. The difference is that $H_3$ involves temporal information and event-level social relation, while $H_4$ only has user-level social information. Besides, we use these two parts to make recommendations separately and show the predicted position of the ground truth event. It is clear that the two parts have different focuses on the event sequence and affect the performance of the model. 
	
	\subsection{Model Complexity}
	We evaluate the temporal and space complexity of STEN by recording the inference time and the model parameters. We compare STEN with the best-performing baselines from three categories mentioned in section 4.2 and results are reported in Table \ref{tab:rq6}. As online recommendation services usually have high requirements on latency, the computational cost during inference is more important than that of the training stage. Compared with the baselines, the increased inference time of STEN is $11\%$ on average, which denotes the efficiency of our model. Besides, we use the number of model parameters to measure the space complexity of models. Despite STEN has performed fine-grained modeling at the event level, it gets the $10\%$ increase on average, within an acceptable range. Compared with the embeddings of both users and items, the parameters in heterogeneous graph component and temporal excitation networks only occupy a small proportion.
	
	\begin{table}[h]	
		\setlength{\abovecaptionskip}{-0cm}
		\caption{Comparsion of model complexity on Ciao dataset.}
		\vspace{0.2cm}
		\centering
		\label{tab:rq6}
		\renewcommand{\arraystretch}{1.2}
		
		\setlength{\tabcolsep}{1.5mm}{
			\resizebox{0.4\textwidth}{!}{
				\begin{tabular}{c|cc|cc}
					\toprule[1pt]
					Method &\tabincell{c}{Inference \\Time }& INCR.&Parameters&INCR.  \\
					\cline{2-5}	
					\hline
					Bert4Rec &$\sim$3.82$s$ &+6.80\%	&$\sim$3.65$\times 10^6$	&+7.74\%\\
					Diffnet	 &$\sim$3.35$s$	&+21.79\%	&$\sim$3.59$\times 10^6$	&+9.53\%\\
					TiSASRec &$\sim$3.91$s$	&+4.45\%	&$\sim$3.48$\times 10^6$	&+12.74\%\\
					STEN	 &$\sim$4.08$s$	&-	&$\sim$3.93$\times 10^6$	&-\\
					\bottomrule[1pt]
				\end{tabular}
			}
		}
	\vspace{-0.5cm}
	\end{table}

	\section{RELATED WORK}
	In this section, we briefly review the related research work from three aspects: social recommendation, general sequential recommendation, and time-aware sequential recommendation.
	
	\textbf{Social Recommendation.} Leveraging social information has played an essential role in alleviating the problem of data sparsity and enhancing recommendation. Kautz et al. \cite{kautz1997referral} import social networks into traditional method collaborative filtering for item recommendation. Ma et al. \cite{ma2008sorec} propose SoRec that incorporates trust network into user tastes analysis through probabilistic matrix factorization. In addition, factor-based random walk is combined with a coupled latent factor model in\cite{yang2011like} to deal with both social connections and interest interactions. Besides, Zhao et al. \cite{zhao2014leveraging} extend the framework of social networks with collaborative filtering to the one-class recommendation. With the development of neural networks, some research works\cite{fan2018deep,wen2018network} utilize deep neural networks to extract information of social networks and learn the complex latent representation. ARSE\cite{sun2018attentive} combines attention mechanism with social LSTM to leverage the temporal information to enhance social recommendation. GraphRec\cite{fan2019graph} utilizes GNNs to model user-user and user-item graphs and heterogeneous strengths in social recommendation. To better simulate the recursive diffusion in social networks, Diffnet\cite{wu2019neural} is designed with an influence propagation structure to model users' latent embedding. To learn latent features from both neighbors and related items, a dual graph attention network model named DANSER\cite{wu2019dual} is proposed and it solves the problem of diversity of neighbor's influence. DGRec\cite{song2019session} utilizes RNNs to model users' dynamic interests and generates the user representation firstly and then combines it with their social relationship through GNNs. 
	
	\textbf{General Sequential Recommendation.} Sequential recommendation has attracted a lot of attention from both academia and industry. The key of sequential recommendation is to extract the hidden information from the user behavior sequence. Rendle et al. \cite{rendle2010factorizing} propose FPMC, which imports Markov Chain (MC) to matrix factorization for user state transition. In addition, RNN and its variants, \textit{e.g.}, Long Short-Term Memory (LSTM) \cite{hochreiter1997long} and Gated Recurrent Unit (GRU) \cite{cho2014learning} are used to model the sequential behaviors of user for their latent features in many related studies\cite{DBLP:journals/corr/HidasiKBT15, wu2017recurrent,yu2016dynamic}. However, RNN-based models usually suffer from low computational efficiency. CNN-based model Caser\cite{2018Personalized} is proposed to use a bi-direction convolution filter to learn sequential patterns. To improve the ability of modeling long sequence, attention-based method Transformer\cite{NIPS2017_7181} and BERT\cite{devlin2018bert} are utilized by SASRec\cite{kang2018self}and BERT4Rec  \cite{sun2019bert4rec} to capture sequential information, respectively. Wu et al. propose SRGNN\cite{wu2019session} that utilizes Gated graph neural network\cite{li2015gated} to capture structure information from sequential data.
	
	\textbf{Time-aware Sequential Recommendation.}
	To take the time interval of user actions into consideration, there also develops many time-sensitive recommendation methods. Time-LSTM\cite{zhu2017next} captures the user's long short-term dynamic interests through the specific time gates. Based on SASRec\cite{kang2018self},  TiSASRec\cite{li2020time} take the time interval between different actions into consideration and model different time intervals as relations between any two interactions.	Besides, another way for the time-sensitive method is to formulate the sequence as a temporal point process. LSHP\cite{cai2018modeling} models the long-term and short-term dependency between users' online interactive behaviors through a Hawkes Process. With the development of neural point process, \textit{e.g.},  neural Hawkes Process\cite{NIPS2017_7252}, CTRec\cite{bai2019ctrec} utilizes a convolutional LSTM and attention mechanism to capture short-term and long-term interests, respectively, based on a demand-aware Hawkes Process. However, RNN-based point process models have limitations in both capturing long-term dependency and computational efficiency. Hawkes process is combined with self-attention mechanism in\cite{zuo2020transformer} to alleviate the limitations. The above time-aware studies leverage the temporal information to model the behavior sequence of individual users. By contrast, our model STEN aims to leverage the temporal information not only to model the user's dynamic interests but also to capture the relation among users through their temporal mutual dependency.

	\section{CONCLUSION AND FUTURE WORK}
In this paper, we propose a novel time-aware sequential recommendation model STEN, to model the fine-grained temporal behavior sequences of users with social information. STEN establishes a social heterogeneous graph to learn the node representations of users and items. In order to model the time-aware behavior sequence, we introduce two classical temporal point processes as prototypes to enhance STEN. STEN models the impact of friends behavior sequences and user's behavior sequence in a direct paradigm by mutually exciting temporal network and self-exciting temporal network. Compared with the previous methods, STEN could model the temporal relationships between cross-user events. As a future plan, we will improve the current model to be type-aware to better model the influence of different types of events and to expand its applicable scenario.

\newpage
	\bibliographystyle{ACM-Reference-Format}
	\bibliography{refer}
	
	\appendix
	%
	%
	%
	%
	%
	%
	%
	%
	
\end{document}